\documentclass[smallextended]{svjour3}       % onecolumn (second format)
\usepackage[ruled,vlined]{algorithm2e}
\usepackage{multirow}
\usepackage[demo]{graphicx}
\usepackage{caption}
\usepackage{subcaption}
\usepackage{float}
\usepackage{amsmath}
\usepackage{booktabs}
\usepackage{multirow}
\usepackage{amsmath}
\usepackage{esvect}
\usepackage[title]{appendix}
\usepackage{hyperref}
\usepackage{dirtytalk}
\usepackage{float}
\hypersetup{
    colorlinks=true,
    linkcolor=blue,
    filecolor=magenta,      
    urlcolor=cyan,
}

\smartqed  % flush right qed marks, e.g. at end of proof
\journalname{Personal and Ubiquitous Computing}
\begin{document}

\title{Diversification in Session-based News Recommender Systems
}
% \subtitle{Do you have a subtitle?\\ If so, write it here}


\author{Alireza Gharahighehi         \and
        Celine Vens %etc.
}


\institute{A. Gharahighehi \and C. Vens \at
              Itec, imec research group at KU Leuven, Kortrijk, Belgium \and KU Leuven, Campus KULAK, Department of Public Health and Primary Care, Kortrijk, Belgium\\
            %   Tel.: +123-45-678910\\
            %   Fax: +123-45-678910\\
              \email{\{alireza.gharahighehi, celine.vens\}@kuleuven.be}          %  \\
            %   \emph{Present address:} of F. Author  %  if needed
%           C. Vens \at
%               second address
}

\date{Received: date / Accepted: date}

\maketitle

\begin{abstract}
Recommender systems are widely applied in digital platforms such as news websites to personalize services based on user preferences. In news websites most of users are anonymous and the only available data is sequences of items in anonymous sessions. Due to this, typical collaborative filtering methods, which are highly applied in many applications, are not effective in news recommendations. In this context, session-based recommenders are able to recommend next items given the sequence of previous items in the active session. Neighborhood-based session-based recommenders has been shown to be highly effective compared to more sophisticated approaches. In this study we propose scenarios to make these session-based recommender systems diversity-aware and to address the filter bubble phenomenon. The filter bubble phenomenon is a common concern in news recommendation systems and it occurs when the system narrows the information and deprives users of diverse information. The results of applying the proposed scenarios show that these diversification scenarios improve the diversity measures in these session-based recommender systems based on four news datasets.
\end{abstract}

\section{Introduction}
\label{sec1}
Nowadays recommender systems (RSs) are applied in almost every digital platform. These platforms try to adapt their services based on user needs in order to increase user satisfaction. RSs infer user needs and preferences using the previous interactions and activities of the user in the platform. In news aggregator websites, users are usually anonymous and therefore their profiles and long-term interaction histories are not available. In this situation the only available information is the sequence of interactions in the active session of the (anonymous) user. Moreover, the news domain is highly dynamic and the set of available news articles for recommendation changes rapidly. Therefore, a news RS should focus on these characteristics to capture recent trends and anonymous users' short-term preferences~\cite{jugovac2018streamingrec,karimi2018news,gharahighehi2019news}.

Session-based Recommender Systems (SBRSs) are applied, when the user's long-term history is not available and the item set is highly dynamic. SBRSs are meant to recommend the next items given the sequence of visited items in the active session of an anonymous user. SBRSs use the collaborative and sequential information from previous sessions of anonymous users to rank and recommend candidate items for an active session. These methods are applied in many applications such as news recommendation~\cite{gabriel2019contextual}, music recommendation and next basket prediction in e-commerce~\cite{jannach2017recurrent}. 

RSs are mostly designed to generate accurate recommendations based on previous interactions. As they are primarily optimized based on predictive accuracy they can narrow the scope of users' recommendations and tighten the filter bubble around the user. The filter bubble phenomenon occurs when users are exposed to similar topics and content and consequently are isolated from diverse viewpoints and content~\cite{nguyen2014exploring}. Pariser~\cite{pariser2011filter} describes a filter bubble as:
\textit{\say{The new generation of Internet filters ... are prediction engines, constantly creating and refining a theory of who you are and what you’ll do and want next. Together, these engines create a unique universe of information for each of us—what I’ve come to call a filter bubble—which fundamentally alters the way we encounter ideas and information.}
}

In news aggregator websites, in addition to the filter bubble phenomenon, focusing only on accuracy can form eco-chambers and boost polarization, radicalization, and fragmentation among users~\cite{helberger2019democratic}.
To mitigate these issues, diversity should be considered in the recommendation lists to avoid recommending redundant items to the users and also to broaden users' horizons.

In this paper we propose diversification approaches for four state-of-art neighborhood-based SBRSs using news article metadata. To the best of our knowledge, most current SBRS methods only focus on providing accurate predictions, ignoring diversity of recommendation lists. In particular, we propose two simple yet effective methods to manipulate the candidate item selection or the neighbor selection, resulting in more diverse recommendation lists. We also study their combined effect. Our main question of interest is investigating the accuracy/diversity trade-off, i.e., quantifying the accuracy loss as a cost for the introduced diversity. For this purpose, we use an evaluation measure that takes both goals into account, apart from the standard accuracy or diversity measures. This is an extension of our previous study~\cite{gharahighehi2020news} where we introduced diversity in the session-based k nearest neighbor (SKNN) method~\cite{jannach2017recurrent}, which is a neighborhood-based SBRS. Our extension in this paper is fourfold:
\begin{itemize}
    \item First, instead of one model (SKNN), we generalize the diversification approaches to several neighborhood-based SBRSs, namely, SKNN~\cite{jannach2017recurrent}, vector multiplication SKNN (VSKNN)~\cite{ludewig2019performance}, STAN~\cite{garg2019sequence} and VSTAN~\cite{ludewig2020empirical}.
    \item Second, we provide a more comprehensive overview of related SBRS studies and diversification approaches.
    \item Third, we compare the results of the proposed approaches with the maximal marginal relevance (MMR) re-ranking approach.
    \item Finally, we extend the evaluation procedure by adding one more news dataset and an additional evaluation measure (topics coverage) to assess the filter bubble issue.
\end{itemize}

In the following, related studies about SBRSs and diversity  are presented in Section~\ref{sec2}. Next, in Section~\ref{sec3}, the proposed scenarios to diversify recommendations of neighborhood-based SBRSs are explained. In Section~\ref{sec4}, four news datasets are described and the experimental setup in designing and testing the proposed scenarios are discussed. Next, the obtained results of applying these proposed scenarios are presented and discussed in Section~\ref{sec5}. Finally, we conclude and outline some future research directions in Section~\ref{sec6}.   

\section{Related work}
\label{sec2}
\subsection{Session-based recommender systems}

SBRSs are studied and applied in many applications such as news, music and e-commerce~\cite{ludewig2020empirical}. In these applications usually the item set is highly dynamic and an anonymous user can receive services without having an explicit profile. Recent state-of-art SBRSs can be categorized to neighborhood- and neural network-based methods.

Neighborhood-based SBRSs form neighborhoods based on the training sessions. Item-based k nearest neighbors (IKNN)~\cite{linden2003amazon} is a naive method that has been traditionally considered as an SBRS baseline where the last item of the active session is used to find neighbors. A generalization of IKNN is session-based KNN (SKNN)~\cite{jannach2017recurrent} that uses all of the items in the active session to find neighbors. Ludewig et al.~\cite{ludewig2019performance} proposed an extension of SKNN that gives more weights to recent events in the session (Vector Multiplication SKNN). Garg et al.~\cite{garg2019sequence} proposed a sequence- and time-aware neighborhood SBRS (STAN) that considers recency and position of the items in SKNN. Recently, Ludewig et al.~\cite{ludewig2020empirical} proposed VSTAN, which is an extension of STAN where they add the sequential information and weighting scheme of Vector Multiplication SKNN (VSKNN) to STAN. We discuss SKNN, VSKNN, STAN and VSTAN in more detail in Section~\ref{sec3_2_1}.

Neural network-based SBRSs have mostly used the recurrent neural network (RNN) structure. GRU4REC~\cite{hidasi2015session} is an RNN-based SBRS that uses Gated Recurrent Units (GRU) to generate recommendations for an active session. There are extensions of GRU4REC where the loss function and sampling approach have been changed to enhance the
performance~\cite{hidasi2018recurrent,hidasi2016parallel}. Li et al.~\cite{li2017neural} proposed another RNN-based SBRS that uses attention mechanism to predict the next item of an active session. Their proposed model (NARM) captures the main purpose of the session to generate recommendations. Recently, Wu et al.~\cite{wu2019session} modeled the current and the global preference of the session with a graph neural network (SR-GNN) SBRS. In this model the sequences of items in the sessions are represented as graph structured data.     

Numerous previous studies on SBRSs~\cite{ludewig2018evaluation,jannach2017recurrent,ludewig2020empirical,ludewig2019performance,kouki2020lab,jugovac2018streamingrec} have shown that neighborhood-based SBRSs surprisingly can outperform recent complex neural network-based SBRSs in both accuracy and computational cost. More specifically,~\cite{ludewig2020empirical} concluded that despite the recent surge of the newly proposed neural network-based SBRSs, in most cases, they are unable to outperform much simpler methods such as neighborhood-based SBRSs. For instance, they showed that VSTAN performs better compared to the recently proposed SR-GNN w.r.t. accuracy, while the training time of VSTAN is 10,000 times lower than SR-GNN.

\subsection{Diversity in recommender systems}
The concept of diversity was first introduced in the information retrieval community. A diversified list is more likely to contain the user's actual search intent~\cite{kaminskas2016diversity}. In recommender systems diversification is applied to provide a wider range of content and therefore to address the filter bubble phenomenon. To measure the diversity of a recommendation list, a common evaluation measure is the average pair-wise distance between items in the ranked list~\cite{smyth2001similarity}. This measure is called intra-list diversity (ILD) and a high value in ILD means the recommended list contains items with a broad range of content. In recommender systems accuracy and ranking play important roles. Vargas and Castells~\cite{vargas2011rank} introduced a rank and relevance sensitive intra-list diversity measure (RR-ILD) that shows to what extent the recommender can diversify the list and preserve the relevant items in the high ranks. It is worth mentioning that there are some other measures related to diversity such as aggregate diversity, calibration and fairness. Aggregate diversity or item coverage measures the portion of available items in the catalog that are included in the recommendation lists~\cite{adomavicius2011maximizing}. A calibrated recommendation list reflects the different areas of interest of a user with their proportions in the user history~\cite{steck2018calibrated}. Fairness measures the degree of parity of recommendations for individuals or groups of users based on a sensitive factor such as gender, ethnicity or age~\cite{dwork2012fairness}.

Generally there are two diversification approaches in recommendation systems: re-ranking and diversity modeling. Re-ranking approaches such as~\cite{kelly2006enhancing,vargas2011intent,jambor2010optimizing,jugovac2017efficient} are post-processing methods that reorder the initial ranked list generated by a baseline recommender. Jugovac et al.~\cite{jugovac2017efficient} proposed a re-ranking approach that can consider multiple quality factors, such as accuracy and diversity, in its re-ranking and adjusts the initial recommendation lists based on tendencies of the individual users. While these methods are able to increase diversity, they need additional post-processing steps that can be computationally demanding when the initial list is long. On the other hand, diversity modeling methods such as~\cite{said2013user,shi2012adaptive,su2013set} adapt the main recommender method to make it diversity-aware based on items metadata. Each of these methods can only be applied on the specific recommender method (for instance BPR) that is used as the main RS and therefore not applicable in other types of RS such as SBRSs.

Although diversity has been vastly studied in user-based recommendation systems, it has received very limited attention in SBRSs. In~\cite{gharahighehi2021personalizing} we proposed a strategy for SKNN to decide between diversification and hybridization for music recommendations. We showed that by hybridizing, i.e. showing more similar content, for the focused sessions and diversifying for the broader sessions, the overall performance in accuracy and diversity could be enhanced in the context of music recommendation. 
While this idea of personalizing diversification is valid for music recommendation, it should not be applied in any news recommender system. In order to avoid possible undesirable side effects of algorithmic filtering such as filter bubbles, fragmentation and radicalization~\cite{helberger2019democratic}, a user should be exposed to a diverse range of ideas and  content. Therefore, a news recommender  should broaden the range of information even for focused sessions. For this reason, in the current article, we show how the diversification approaches are able to extend the range of news topics in recommendation lists, by maximizing the overall RR-ILD.
Furthermore, unlike music datasets, news datasets normally contain several sources of information such as text, summary, named entities and tags, which can be represented using article embeddings and exploiting when constructing neighborhoods. To fully investigate and compare the potential of neighborhood-based methods to increase diversity, we apply the diversity enhancing techniques also to three more recent methods, besides SKNN, that were discussed before (VSKNN, STAN and VSTAN).

\section{Methodology}
\label{sec3}

In this section we explain how we make neighborhood-based SBRSs diversity-aware. We first explain the background related to these SBRSs, namely, SKNN \cite{jannach2017recurrent}, VSKNN~\cite{ludewig2019performance}, STAN~\cite{garg2019sequence} and VSTAN~\cite{ludewig2020empirical} in Section~\ref{sec3_2_1} and then propose our approaches to make them diversity-aware in Section~\ref{sec3_2_2}. As mentioned above, these SBRSs are very promising in terms of accuracy and computational cost~\cite{ludewig2018evaluation,jannach2017recurrent,jugovac2018streamingrec,ludewig2020empirical,ludewig2019performance,kouki2020lab}, compared to more complex types of SBRS.

\subsection{Background}
\label{sec3_2_1}

SKNN~\cite{jannach2017recurrent} is a memory-based SBRS that forms a neighborhood to provide recommendations. It uses the items in the active session to select the nearest neighbor sessions and to predict the next items in the active session. To predict the score of a candidate item, SKNN uses the similarity of the item set in the neighbor sessions that include the candidate item, with the item set in the active session. This score is calculated using Eq.~\ref{equ:5}:

\begin{equation}
\label{equ:5}
\hat{r}_{_{SKNN}}(i,s) = \sum_{n\in N_{s}} w_{s,n}\times 1_{n}(i)
\end{equation}

\noindent where \(\hat{r}_{_{SKNN}}(i,s)\) is the predicted score for active session \(s\) and candidate item \(i\), \(w_{s,n}\) is the similarity between session \(s\) and session \(n\), \(1_{n}(i)\) is an indicator function that verifies whether item \(i\) exists in session \(n\) and \(N_{s}\) is the set of neighbor sessions for session \(s\). To calculate the similarity between two sessions one can use \textit{cosine} distance measure:

\begin{equation}
\label{equ:sknn_w}
w_{s,n} = \dfrac{\vv{s} \cdot \vv{n}}{\lVert \vv{s} \rVert \times \lVert \vv{n} \rVert}
\end{equation}

\noindent where \(\vv{s}\) and \(\vv{n}\) are the binary vector representations for session \(s\) and \(n\) over the item set in training sessions. SKNN does not consider the order of items in sessions and time in its predictions.

VSKNN~\cite{ludewig2019performance} is an extension of SKNN that puts more focus on the recent items in the active session. In this method, instead of binary values, real values are assigned to the items of the session based on their order. The last item of the active session gets the value 1 and the values of the less recent items in the vector are decayed based on a decay function. The choice of the decay function is a hyperparameter. Ludewig et al.~\cite{ludewig2019performance} proposed linear, logarithmic, quadratic and inverse decay functions. For instance, using the inverse decay function, if the active session contains three items then the last item in the session gets the value one, the next one gets the value \(\frac{1}{2}\) and the fist item of the session gets the value \(\frac{1}{3}\) in the session vector. After constructing this vector with real values using one of these decay functions, \(w_{s,n}\) is defined as the similarity between the binary vector of the neighbor session and this real-valued vector of the active session. 

STAN~\cite{garg2019sequence} is another extension of SKNN that considers three additional components in SKNN to calculate relevance scores: (1) position of items in the active session, (2) recency of the neighbor session and (3) position of the candidate item in the neighbor session. They proposed three decay functions to include these three components in SKNN. To focus more on the more recent items in the active session (the first additional component), real values are calculated for items in the active session using the following exponential decay function:

\begin{equation}
\label{equ:vkss_w}
s_{i} = exp(\dfrac{p(s,i)-l(s)}{\lambda_{1}})
\end{equation}

\noindent where \(i\) is an item in the active session \(s\), \(p(s,i)\) is the position of item \(i\) in session \(s\), \(l(s)\) is the position of the last item of session \(s\) and \(\lambda_{1}\) is a hyperparameter (e.g. if \(i\) is the third item of a session with length 4, \(p(s,i)=3\) and \(l(s)=4\)). The similarity of this real-valued vector \(\vv{s}\) and the binary vector of neighbor session \(n\) can be calculated using Eq.~\ref{equ:sknn_w}. This is similar to VSKNN but in STAN the exponential decay function (Eq.~\ref{equ:vkss_w}) is used to calculate the real-values in vector \(\vv{s}\).

To give a higher weight to the more recent neighbor session w.r.t. active session (the  second  additional  component), they used the following decay function:

\begin{equation}
\label{equ:vkss_w}
w_{t}(s,n) = exp(- \dfrac{t(s)-t(n)}{\lambda_{2}})
\end{equation}

\noindent where \(t(s)\) and \(t(n)\) are the timestamps of the last event/item of the active session \(s\) and the neighbor session \(n\) (\(t(s)>t(n)\)) and \(\lambda_{2}\) is a hyperparameter. Finally to consider the position of the candidate item in the neighbor session (the  third  additional  component), the following decay function is proposed:

\begin{equation}
\label{equ:vkss_w}
w_{n}(i,n) = exp(- \dfrac{\lvert p(i,n)-p(i^{\ast},n)\rvert}{\lambda_{3}})\times I_{n}(i) \times I_{n \cap s}(i^{\ast})
\end{equation}

\noindent where \(i^{\ast}\) is the most recent item that occurred in the active session \(s\) and that also occurred in the neighbor session \(n\) and \(\lambda_{3}\) is a hyperparameter. The final predicted relevance score in STAN is:

\begin{equation}
\label{equ:stan}
\hat{r}_{_{STAN}}(i,s) = \sum_{n\in N_{s}} w_{s,n} \times w_{t}(s,n) \times w_{n}(i,n) \times1_{n}(i)
\end{equation}

VSTAN~\cite{ludewig2020empirical} is the most recent extension of SKNN that has two additional components compared to STAN: (1) position of the last item in the active session that also exists in the neighbor session (Eq.~\ref{equ:wsn}) and (2) an \textit{idf}\footnote{Inverse Document Frequency} weighting scheme (Eq.\ref{equ:idf}) that gives lower weights to frequent candidate items:

\begin{eqnarray}
\label{equ:wsn}
& & w_{s}(n,s) = exp(\dfrac{p(i^{\ast},s)-l(s)}{\lambda_{4}}) \times I_{n \cap s}(i^{\ast})\\
& & idf_{i} = log(\dfrac{\lvert S\rvert }{\lvert\{s \in S: i \in s \}\rvert})\times \lambda_{idf}
\label{equ:idf}
\end{eqnarray}

\noindent where \(w_{s}(n,s)\) is the position weight of the last item (\(i^{\ast}\)) in active session \(s\) that also exists in neighbor session \(n\), \(idf_{i}\) is the \(idf\) weight of candidate item \(i\), \(S\) is the set of training sessions and \(\lambda_{4}\) and \(\lambda_{idf}\) are hyperparameters. The relevance score using VSTAN is calculated by:

\begin{equation}
\label{equ:vstan}
\hat{r}_{_{VSTAN}}(i,s) = \sum_{n\in N_{s}} w_{s,n} \times w_{t}(n,s) \times w_{n}(i,n) \times w_{s}(n,s) \times idf_{i} \times 1_{n}(i)
\end{equation}

\subsection{Diversification approaches}
\label{sec3_2_2}
To make these neighborhood-based SBRSs (SKNN, VSKNN, STAN and VSTAN) diversity-aware we propose three diversification approaches: (1) \textit{diverse neighbor}, (2) \textit{diverse candidate item} and (3) their combination. We need a content representation for news articles and a distance measure to apply these approaches. We use article embeddings, which are described in the next section, as the content representation, and \textit{cosine} similarity measure in this study.

\paragraph{Diverse neighbor:} In this approach we consider higher weights for the neighbor sessions that have more content diversity. Diversity of a neighbor session is the average content dissimilarity of pairs of items in the session. Neighbors with higher average content dissimilarity are more probable to bring more diverse recommendations. The diversity of a neighbor session can be calculated as following:

\begin{equation}
\label{equ:dn}
    d_{n} = \dfrac{\sum_{i\in n}\sum_{j\in n \texttt{\char`\\} \{i\}}dist_{c}(i,j)}{|n|(|n|-1)}\\
\end{equation}

\paragraph{Diverse candidate item:} In this approach we consider a higher diversity weight for a candidate item with higher average dissimilarity with items of the active session. This weight is calculated using the embedding of the candidate item and the embeddings of the articles in the active session based on the following equation:

\begin{equation}
\label{equ:6}
    d_{i,s} = \dfrac{\sum_{j\in s}dist(i,j)}{|s|}
\end{equation}

\noindent where \(d_{i,s}\) is the diversity weight of candidate item \(i\) and active session \(s\) and \(dist(i,j)\) is the dissimilarity between embeddings of item \(i\) and \(j\). These two proposed diversification approaches can be simultaneously applied to enhance diversity. 

To make SKNN diversity-aware we apply these two introduced diversity weights: 

\begin{equation}
\label{equ:d_sknn}
\hat{r}_{_{d-SKNN}}(i,s) = d_{i,s}\sum_{n\in N_{s}} w_{s,n}\times d_{n}\times 1_{n}(i)\\        
\end{equation}

\noindent where \(d_{n}\) is the diversity of session \(n\) and \(d_{i,s}\) is the average content dissimilarity of item \(i\) and the items in session \(s\). Similar weights (\(d_{n}\) and \(d_{i,s}\)) are added to VSKNN, STAN and VSTAN to make them diversity-aware.

To compare the results of the proposed approaches with a diversification baseline, we use the Maximal Marginal Relevance (MMR) re-ranking approach~\cite{kaminskas2016diversity}. In this approach multiple performance criteria (here accuracy and diversity) are used to re-rank items of an initial recommendation list.

\section{Dataset and Experimental Setup}
\label{sec4}
We evaluate three diversity boosting approaches, namely, \textit{diverse neighbor}, \textit{diverse candidate item} and the combination of both with the original methods. We use four news datasets, namely \textit{Adressa}~\footnote{http://reclab.idi.ntnu.no/dataset}~\cite{gulla2017adressa}, \textit{Globo.com}~\footnote{https://www.kaggle.com/gspmoreira/news-portal-user-interactions-by-globocom}~\cite{gabriel2019contextual}, \textit{Kwestie} and \textit{Roularta}\footnote{The pre-processed version of the datasets are available upon request. The last two datasets were obtained from Roularta Media Group, a Belgian multimedia group.}, which are described in Table~\ref{tab:0}, to evaluate the performance of these approaches. To calculate content dissimilarity, as explained in the previous section, we should form content representations for news articles. The CNN-based deep neural network approach proposed by~\cite{gabriel2019contextual} is used to generate article embeddings for news articles of these datasets based on article title, summary, full text and tags. 

\begin{table*}
\centering
  \caption{Datasets descriptions}
    \label{tab:0} 
   
  \begin{tabular}{lcccc}
    \toprule
    & Roularta&Kwestie&Globo.com&Adressa\\
    \midrule
    \# Sessions &324,505&297,342&1,048,389&5,607,303\\
       \# Interactions&5,529,007&922,680&2,986,226&8,894,565\\
   \# Items&29,641&22,465&45,559&27,805\\
%   \# Topics&169&51&101&194\\
   Timespan&26 days&28 days&16 days& 27 days\\
   Language&Dutch and French&Dutch&Portuguese&Norwegian\\
  \bottomrule
\end{tabular}
\end{table*}

We compare performance of the SBRSs explained in the previous section (SKNN, VSKNN, STAN and VSTAN) with the proposed diversification approaches, \textit{diverse candidate item} (\_I), \textit{diverse neighbor} (\_D), the combined  approaches (\_ID) and the MMR re-ranking approach (\_Re) based on five performance measures, namely precision (\textit{P@k}), recall (\textit{R@k}), expected intra-list diversity (\textit{ILD@k}), rank and relevance sensitive expected intra-list diversity (\textit{RR-ILD@k})~\cite{vargas2011rank} and covered topics (\textit{CT@k}). \textit{P@k} and \textit{R@k} are standard information retrieval accuracy measures that evaluate the model in predicting the relevant items in the top k recommendations. \textit{ILD@k} is the average content dissimilarity between pairs of items in top k recommendation and \textit{RR-ILD@k} is another diversity measure, that considers the ranks and relevance of top k recommendation in calculating diversity. \textit{CT@k} is the average number of unique topics/keywords in recommendation lists. \textit{CT@k} is used to evaluate the merits of the diversification approaches in expanding the range of recommended content to the users by enhancing the number of news topics in the recommendation lists and consequently addressing the filter bubble issue. A recommendation list with more unique topics is less probable to tighten the filter bubble around the user.

To form the train and test sets, we use the approach by~\cite{ludewig2018evaluation}. In this approach, the datasets are split into five partitions with the same duration. The sessions in the last day of each partition are considered as test sessions, and the sessions from the other days of the partition as training sessions. In these test sessions the last two items are regarded as the test items. The accuracy measures (\textit{P@k} and \textit{R@k}) are calculated based on the ability of the model in predicting these test items in the test sessions. The reported performance in the next section is the average performance of the model over these five partitions. 

As mentioned in the previous section, there are some hyperparameters to be tuned. We consider the last day of the first training set as the validation set and tune the hyperparameters based on \textit{P@10}. The final values of the tuned hyperparameters are reported in Table~\ref{tab:01}. In order to decide which diversification approach should be applied, one can select the approach based on the performance  in the validation set. In our experiments, we select the approach that yields the highest \textit{RR-ILD@k} in the validation set, as it  provides a trade-off between accuracy and diversity. The selected approach for each pair of method-dataset is indicated in Table~\ref{tab:01}.

\begin{table*}
\centering
  \caption{The final tuned values for the hyperparameters of the neighborhood-based SBRSs} 
    \label{tab:01} 
    \scalebox{0.90}{%
  \begin{tabular}{lp{0.14\textwidth}ccccc}
  \hline\noalign{\smallskip}
    Method&Hyper-parameters&Range& \textbf{Adressa}&\textbf{Globo}&\textbf{Kwestie}&\textbf{Roularta}\\
    \noalign{\smallskip}\hline\noalign{\smallskip}
    % \multirow{}{*}{SR}
    % &Max \# events between items& [1,20]& 13 & 11 & 20 & 12\\
    %     % &sim & div  & same & same & same\\
    %   \noalign{\smallskip}\hline 
   \multirow{3}{*}{SKNN}
   &sample size& [500,3000]& 2500 & 700 & 500 & 500\\
    &\# Neighbors&[50,500] & 300 & 500 & 300 & 200\\
    &approach&[\_I,\_D,\_ID]&\_D&\_ID&\_I&\_ID\\
    % & similarity& & cosine & jaccard & cosine&cosine\\
    %   \noalign{\smallskip}\hline 
    % \multirow{3}{*}{SSKNN}&Sample size& 500 & 100 & 100 & 50\\
    % &\# Neighbors & 300 & 100 & 100 & 50\\
    % &\# Weighting & quad & 100 & 100 & 50\\
       \noalign{\smallskip}\hline 
    \multirow{4}{*}{VSKNN}&sample size&[500,3000]& 2500 & 500 & 2500 & 500\\
    &\# Neighbors&[50,500] & 300 & 100 & 300 & 50\\
    &\# Weighting& [inverse,linear,quad,log]& inverse & log & log & quad\\
    &approach&[\_I,\_D,\_ID]&\_D&\_ID&\_I&\_ID\\
    %  &\# Weighting score & quad & quad & quad & same\\
    % &\# IDF weighting & - & - & - & -\\
       \noalign{\smallskip}\hline 
    \multirow{5}{*}{STAN}&sample size&[500,3000] & 1000 & 500 & 2500 & 1000\\
    &\# Neighbors &[50,500]& 100 & 50 & 100 & 50\\
    & $\lambda_{spw}$& [0.1,5]& 0.80 & 1.15 & 5.00 & 2.55\\
    & $\lambda_{snh}$&[1,5] & 1.29 & 1.00 & 4.42 & 1.86\\
    & $\lambda_{inh}$&[0.5,5] & 5.00 & 2.75 & 5.00 & 4.68\\
    &approach&[\_I,\_D,\_ID]&\_D&\_I&\_I&\_ID\\
       \noalign{\smallskip}\hline 
     \multirow{7}{*}{VSTAN}
     &sample size& [500,3000]& 1000 & 700 & 2500 & 500\\
    &\# Neighbors& [50,500]& 100 & 100 & 300& 50\\
   &$\lambda_{spw}$&[0.1,5] & 1.50 & 1.85 & 3.95 & 3.60\\
   &$\lambda_{snh}$&[1,5] & 1.57 & 3.57 & 2.71 & 2.14\\
   &$\lambda_{inh}$&[0.5,5] & 3.39 & 2.11 & 4.68 & 3.71\\
   &$\lambda_{ipw}$&[0.1,5] &3.95 & 1.50 & 3.25 & 3.25\\
    &$\lambda_{idf}$& [0.1,5]&0.10 & 4.3 & 1.50 & 0.10\\
    &approach&[\_I,\_D,\_ID]&\_D&\_I&\_I&\_ID\\
   \noalign{\smallskip}\hline 
\end{tabular}}
\end{table*}

\section{Result and discussion}
\label{sec5}
The results of applying the proposed approaches\footnote{The source code is available at \url{https://github.com/alirezagharahi/d_SBRS}} for \textit{Adressa} dataset w.r.t. \textit{P@10}, \textit{R@10}, \textit{ILD@10}, \textit{RR-ILD@10} and \textit{CT@10} are reported in Table~\ref{tab:adressa_n}. 
As is shown in Table~\ref{tab:adressa_n}, while \textit{SKNN} is the simplest form of neighborhood-based SBRSs, it has the best accuracy and diversity among these methods. Among the proposed approaches, the \textit{\_ID} approach has the best performance in \textit{ILD@10} but it deteriorates the accuracy significantly. In \textit{SKNN} and \textit{VSKNN} the re-ranking approach performs better w.r.t. \textit{RR-ILD@10} compared to the proposed approaches. On the other hand, in \textit{STAN} and \textit{VSTAN}, the \textit{\_D} approach has the best \textit{RR-ILD@10} and therefore can provide a better trade-off between diversity and accuracy. It also yields the two overall highest \textit{RR-ILD@10} values among all optimized neighborhood based models. 

\begin{table*}[h]
\centering
  \caption{Results of diversity-aware neighborhood-based SBRSs for \textit{Adressa}. Bold values highlight the best performing method and italic values indicate improvement w.r.t.~the base method.}
  \label{tab:adressa_n}
  \begin{tabular}{lccccc}
\hline\noalign{\smallskip}
    Methods&P@10&R@10&ILD@10&RR-ILD@10&CT@10\\
\hline\noalign{\smallskip}
SKNN & 0.0788 & 0.3942 & 0.1849 & 0.0198 & 19.28\\
SKNN\_I & 0.0654 & 0.3272 & \textit{0.2139} & 0.0179 & \textit{19.99}\\
SKNN\_D & 0.0786 & 0.3929 & \textit{0.1927} & \textit{0.0204} & \textit{19.62}\\
SKNN\_ID & 0.0631 & 0.3153 & \textit{\textbf{0.2214}} & 0.0177 & \textit{20.14}\\
SKNN\_Re   & \textit{\textbf{0.0800}} & \textit{\textbf{0.4001}} & \textit{0.2139} & \textit{\textbf{0.0224}} & \textit{\textbf{20.67}} \\
\hline\noalign{\smallskip}
VSKNN & \textbf{0.0780} & \textbf{0.3898} & 0.1848 & 0.0196 & 19.26\\
VSKNN\_I & 0.0540 & 0.2699 & \textit{0.2052} & 0.0143 & \textit{19.77}\\
VSKNN\_D & 0.0768 & 0.3840 & \textit{0.1920} & \textit{0.0202} & \textit{19.79}\\
VSKNN\_ID & 0.0523 & 0.2614 & \textit{0.2114} & 0.0140 & \textit{19.99}\\
VSKNN\_Re  & 0.0760 & 0.3800 & \textit{\textbf{0.2134}} & \textit{\textbf{0.0219}} & \textit{\textbf{20.93}} \\
\hline\noalign{\smallskip}
STAN & \textbf{0.0784} & \textbf{0.3920} & 0.1838 & 0.0196 & 19.91\\
STAN\_I & 0.0572 & 0.2858 & \textit{0.2050} & 0.0151 & \textit{20.53}\\
STAN\_D & 0.0751 & 0.3755 & \textit{0.2244} & \textit{\textbf{0.0230}} & \textit{21.01}\\
STAN\_ID & 0.0497 & 0.2483 & \textit{\textbf{0.2405}} & 0.0153 & \textit{\textbf{21.18}}\\
STAN\_Re   & 0.0778 & 0.3889 & \textit{0.1907} & \textit{0.0200} & \textit{20.26} \\
\hline\noalign{\smallskip}
VSTAN & \textbf{0.0784} & \textbf{0.3920} & 0.1835 & 0.0195 & 19.43\\
VSTAN\_I & 0.0570 & 0.2851 & \textit{0.2059} & 0.0153 & \textit{20.38}\\
VSTAN\_D & 0.0745 & 0.3723 & \textit{0.2265} & \textit{\textbf{0.0231}} & \textit{\textbf{21.01}}\\
VSTAN\_ID & 0.0489 & 0.2444 & \textit{\textbf{0.2447}} & 0.0156 & \textit{20.87}\\
VSTAN\_Re  & 0.0782 & 0.3911 & \textit{0.1890} & \textit{0.0199} & \textit{19.79}\\
  \hline\noalign{\smallskip}
\end{tabular}
\end{table*}

For the \textit{Globo} dataset, the results of applying the proposed approaches are shown in Table~\ref{tab:globo_n}. According to this table, all the proposed approaches enhance diversity and number of unique topics in recommendations with the cost of reduced accuracy. 
In neighborhood-based SBRSs, STAN has the best performance in all measures. The \textit{\_ID} approach has the best \textit{RR-ILD@10} and \textit{CT@10} in \textit{SKNN} and \textit{VSKNN}, but the \textit{\_I} approach has better \textit{RR-ILD@10} in \textit{STAN} and \textit{VSTAN} among the other approaches. The re-ranking approach can not outperform the best performing proposed approach in \textit{RR-ILD@10} and therefore is less effective.

\begin{table*}[h]
\centering
  \caption{Results of diversity-aware neighborhood-based SBRSs for \textit{Globo}. Bold values highlight the best performing method and italic values indicate improvement w.r.t.~the base method.}
  \label{tab:globo_n}
  \begin{tabular}{lcccccc}
\hline\noalign{\smallskip}
    Methods&P@10&R@10&ILD@10&RR-ILD@10&CT@10\\
\hline\noalign{\smallskip}
SKNN & \textbf{0.0444} & \textbf{0.2218} & 0.3382 & 0.0162 & 7.04 \\
SKNN\_I & 0.0417 & 0.2085 & \textit{0.3613} & \textit{0.0191} & \textit{7.59} \\
SKNN\_D & 0.0441 & 0.2206 & \textit{0.3466} & \textit{0.0166} & \textit{7.28} \\
SKNN\_ID & 0.0408 & 0.2039 & \textit{0.3668} & \textit{\textbf{0.0192}} & \textit{\textbf{7.79}} \\
SKNN\_Re   & 0.0422 & 0.2108 & \textit{\textbf{0.3699}} & \textit{0.0173} & \textit{7.75} \\
\hline\noalign{\smallskip}
VSKNN & \textbf{0.0461} & \textbf{0.2304} & 0.3383 & 0.0170 & 7.12 \\
VSKNN\_I & 0.0450 & 0.2250 & \textit{0.3561} & \textit{0.0215} & \textit{7.54} \\
VSKNN\_D & 0.0460 & 0.2302 & \textit{0.3463} & \textit{0.0175} & \textit{7.32} \\
VSKNN\_ID & 0.0442 & 0.2210 & \textit{0.3634} & \textit{\textbf{0.0218}} & \textit{\textbf{7.73}} \\
VSKNN\_Re  & 0.0422 & 0.2110 & \textit{\textbf{0.3696}} & \textit{0.0176} & \textit{7.79} \\
\hline\noalign{\smallskip}
STAN & \textbf{0.0468} & \textbf{0.2341} & 0.3414 & 0.0174 & 7.15 \\
STAN\_I & 0.0454 & 0.2271 & \textit{0.3602} & \textit{\textbf{0.0218}} & \textit{7.73} \\
STAN\_D & 0.0432 & 0.2161 & \textit{0.3619} & \textit{0.0176} & \textit{7.63} \\
STAN\_ID & 0.0413 & 0.2066 & \textit{\textbf{0.3727}} & \textit{0.0206} & \textit{\textbf{8.10}} \\
STAN\_Re   & 0.0457 & 0.2286 & \textit{0.3551} & \textit{0.0178} & \textit{7.46} \\
\hline\noalign{\smallskip}
VSTAN & \textbf{0.0448} & \textbf{0.2241} & 0.3382 & 0.0165 & 7.06 \\
VSTAN\_I & 0.0438 & 0.2189 & \textit{0.3632} & \textit{\textbf{0.0207}} & \textit{7.87} \\
VSTAN\_D & 0.0408 & 0.2041 & \textit{0.3585} & \textit{0.0166} & \textit{7.57} \\
VSTAN\_ID & 0.0378 & 0.1891 & \textit{\textbf{0.3754}} & \textit{0.0193} & \textit{\textbf{8.20}} \\
VSTAN\_Re  & 0.0448 & 0.2241 & \textit{0.3414} & \textit{0.0166} & \textit{7.17} \\
  \hline\noalign{\smallskip}
\end{tabular}
\end{table*}

Table~\ref{tab:kwestie_n} shows the performance of the proposed methods for the \textit{Kwestie} dataset. As is shown in this table,
 
\textit{STAN} has the best accuracy, diversity and \textit{RR-ILD@10}, while \textit{VSTAN} has the best topic coverage compared to the other baselines. All the proposed diversification approaches can improve diversity, \textit{RR-ILD@10} and topic coverage.  In \textit{SKNN} and \textit{VSKNN} the \textit{\_I} approach and the re-ranking approach have almost same performance in \textit{RR-ILD@10}. In STAN and VSTAN the \textit{\_I} approach can effectively improve diversity while slightly reducing accuracy in all methods and therefore has the best \textit{RR-ILD@10} among the diversification approaches.

\begin{table*}[h]
\centering
  \caption{Results of diversity-aware neighborhood-based SBRS for \textit{Kwestie}. Bold values highlight the best performing method and italic values indicate improvement w.r.t.~the base method.}
  \label{tab:kwestie_n}
  \begin{tabular}{lcccccc}
\hline\noalign{\smallskip}
    Methods&P@10&R@10&ILD@10&RR-ILD@10&CT@10\\
\hline\noalign{\smallskip}
SKNN & \textbf{0.0347} & \textbf{0.1735} & 0.1693 & 0.0069 & 3.71 \\
SKNN\_I & 0.0237 & 0.1185 & \textit{0.2978} & \textit{\textbf{0.0091}} & \textit{5.40} \\
SKNN\_D & 0.0309 & 0.1543 & \textit{0.2083} & \textit{0.0077} & \textit{4.07} \\
SKNN\_ID & 0.0195 & 0.0976 & \textit{\textbf{0.3223}} & \textit{0.0090} & \textit{\textbf{5.84}} \\
SKNN\_Re   & 0.0329 & 0.1643 & \textit{0.2308} & \textit{0.0090} & \textit{4.34} \\
\hline\noalign{\smallskip}
VSKNN & \textbf{0.0353} & \textbf{0.1763} & 0.1685 & 0.0066 & 3.69 \\
VSKNN\_I & 0.0252 & 0.1260 & \textit{0.2683} & \textit{0.0087} & \textit{5.03} \\
VSKNN\_D & 0.0312 & 0.1561 & \textit{0.2092} & \textit{0.0077} & \textit{4.07} \\
VSKNN\_ID & 0.0214 & 0.1070 & \textit{\textbf{0.2913}} & \textit{0.0086} & \textit{\textbf{5.34}} \\
VSKNN\_Re  & 0.0321 & 0.1605 & \textit{0.2291} & \textit{\textbf{0.0088}} & \textit{4.30} \\
\hline\noalign{\smallskip}
STAN & \textbf{0.0415} & \textbf{0.2075} & 0.1752 & 0.0084 & 3.80 \\
STAN\_I & 0.0341 & 0.1703 & \textit{0.2679} & \textit{\textbf{0.0113}} & \textit{5.13} \\
STAN\_D & 0.0355 & 0.1776 & \textit{0.2340} & \textit{0.0093} & \textit{4.45} \\
STAN\_ID & 0.0270 & 0.1352 & \textit{\textbf{0.3019}} & \textit{0.0111} & \textit{\textbf{5.74}} \\
STAN\_Re   & 0.0406 & 0.2029 & \textit{0.1876} & \textit{0.0085} & \textit{3.97} \\
\hline\noalign{\smallskip}
VSTAN & \textbf{0.0370} & \textbf{0.1850} & 0.1817 & 0.0077 & 4.09 \\
VSTAN\_I & 0.0277 & 0.1385 & \textit{0.2834} & \textit{\textbf{0.0106}} & \textit{5.64} \\
VSTAN\_D & 0.0327 & 0.1635 & \textit{0.2361} & \textit{0.0092} & \textit{4.67} \\
VSTAN\_ID & 0.0219 & 0.1094 & \textit{\textbf{0.3061}} & \textit{0.0096} & \textit{\textbf{6.10}} \\
VSTAN\_Re  & 0.0370 & 0.1849 & \textit{0.1861} & 0.0077 & \textit{4.15}\\
  \hline\noalign{\smallskip}
\end{tabular}
\end{table*}

For the \textit{Roularta} dataset, the results of applying the proposed approaches are reported in Table~\ref{tab:Roularta_n}. 
According to this table, \textit{STAN} has the best accuracy, diversity and \textit{RR-ILD@10} and \textit{VSTAN} provides recommendations that cover more unique topics compared to the other neighborhood-based baselines. All the proposed approaches improve diversity, \textit{RR-ILD@10} and topic coverage. For \textit{VSTAN} the \_{I} approach is the most effective way to diversify the recommendations, while in the other neighborhood-based approaches the \_{ID} approach is the most effective approach among the proposed diversification approaches. 

\begin{table*}[h]
\centering
  \caption{Results of diversity-aware neighborhood-based SBRS for \textit{Roularta}. Bold values highlight the best performing method and italic values indicate improvement w.r.t.~the base method.}
  \label{tab:Roularta_n}
  \begin{tabular}{lcccccc}
\hline\noalign{\smallskip}
    Methods&P@10&R@10&ILD@10&RR-ILD@10&CT@10\\
\hline\noalign{\smallskip}
SKNN       & \textbf{0.0351} & \textbf{0.1754} & 0.1633 & 0.0063 & 4.34 \\
SKNN\_I    & 0.0279 & 0.1395 & \textit{0.2463} & \textit{0.0078} & \textit{5.16} \\
SKNN\_D   & 0.0348 & 0.1742 & \textit{0.1808} & \textit{0.0067} & \textit{4.60} \\
SKNN\_ID  & 0.0244 & 0.1220 & \textit{\textbf{0.2679}} & \textit{\textbf{0.0079}} & \textit{\textbf{5.42}} \\
SKNN\_Re   & 0.0344 & 0.1720 & \textit{0.1965} & \textit{0.0072} & \textit{4.66} \\
\hline\noalign{\smallskip}
VSKNN      & \textbf{0.0340} & \textbf{0.1701} & 0.1626 & 0.0065 & 4.37 \\
VSKNN\_I   & 0.0283 & 0.1416 & \textit{0.2300} & \textit{0.0088} & \textit{5.04} \\
VSKNN\_D  & 0.0298 & 0.1491 & \textit{0.2106} & \textit{0.0073} & \textit{4.95} \\
VSKNN\_ID  & 0.0257 & 0.1286 & \textit{\textbf{0.2522}} & \textit{\textbf{0.0090}} & \textit{\textbf{5.28}} \\
VSKNN\_Re  & 0.0333 & 0.1667 & 0.\textit{2021} & \textit{0.0076} & \textit{4.84} \\
\hline\noalign{\smallskip}
STAN       & \textbf{0.0462} & \textbf{0.2311} & 0.1660 & 0.0085 & 4.37 \\
STAN\_I    & 0.0419 & 0.2093 & \textit{0.2147} & \textit{0.0118} & \textit{4.93} \\
STAN\_D    & 0.0445 & 0.2225 & \textit{0.1793} & \textit{0.0088} & \textit{4.56} \\
STAN\_ID   & 0.0394 & 0.1972 & \textit{\textbf{0.2339}} & \textit{\textbf{0.0119}} & \textit{\textbf{5.25}} \\
STAN\_re   & 0.0459 & 0.2297 & \textit{0.1701} & \textit{0.0086} & \textit{4.40} \\
\hline\noalign{\smallskip}
VSTAN      & 0.0408 & 0.2041 & 0.1653 & 0.0076 & 4.46 \\
VSTAN\_I   & 0.0335 & 0.1674 & \textit{0.2365} & \textit{\textbf{0.0111}} & \textit{5.44} \\
VSTAN\_D   & 0.0385 & 0.1926 & \textit{0.1910} & \textit{0.0083} & \textit{4.84} \\
VSTAN\_ID  & 0.0281 & 0.1403 & \textit{\textbf{0.2543}} & \textit{0.0099} & \textit{\textbf{5.74}} \\
VSTAN\_re  & \textbf{0.0448} & \textbf{0.2239} & \textit{0.1688} & \textit{0.0083} & \textit{4.44}\\
  \hline\noalign{\smallskip}
\end{tabular}
\end{table*}

To summarize, the recommendation lists have more diversity when more diverse neighbors are selected and when candidate items that have more dissimilarities with the corresponding active sessions are recommended.
In the neighborhood-based SBRSs, the nearest neighbors convey the collaborative information and, according to the results, using more diverse collaborative information gives a better trade-off between diversity and accuracy, i.e., it has always higher \textit{RR-ILD@10} compared to the original methods. The \textit{diverse candidate item} approach is based on content-based information and in all methods and datasets it can bring more diverse recommendations. While all the proposed approaches are able to enhance \textit{ILD@10}, their capabilities should be evaluated based on \textit{RR-ILD@10} as it shows to what extent the trade-off between diversity and accuracy is achieved. These two diversity metrics may have different behavior as they measure the diversity of recommendation lists differently (e.g. in Table~\ref{tab:kwestie_n} the best approach based on \textit{ILD@10} is \_{ID} which is not the case for \textit{RR-ILD@10}). Almost in all cases, at least one of the proposed diversification approaches can enhance \textit{RR-ILD@10} compared to the original neighborhood-based methods and also can outperform the re-ranking approach. The choice of the best and the most effective approach depends on the SBRS method and the dataset. This is in line with the results of~\cite{ludewig2018evaluation,ludewig2020empirical} where the choice of best performing SBRS depends on the dataset. Important to note is that the best performing diversification approaches in \textit{RR-ILD@10} for all methods and datasets are in line with the approaches selected in the validation phase (See Table~\ref{tab:01}). 

To investigate how the proposed approaches address the filter bubble phenomenon, the recommended lists generated by the diversity-aware approaches and the original models are compared w.r.t. the average number of unique topics (\textit{CT@10}) that they recommend to the users. All the proposed approaches enhance \textit{CT@10} in all datasets, which indicates that the diversification approaches broaden the user reading experience and offer a wider range of content by recommending articles from more diverse topics and therefore mitigate the filter bubble issue. The topics in these datasets refer to the main themes (e.g. entertainment, political news, health, ...) that form the content of these news websites. In the Globo, Kwestie and Roularta datasets each article is connected to one topic and therefore the maximum value for \textit{CT@10} in these datasets is 10. In the Adressa dataset multiple topics can be assigned to each article and therefore \textit{CT@10} can be higher than 10. By covering more topics in recommendation lists, users are exposed to news from different themes and therefore have a more diverse reading experience. As is shown in Table~\ref{tab:prearson}, in all datasets the Pearson correlations between diversity (\textit{ILD@10}) and \textit{CT@10} of the methods are strong and statistically significant (p-values are less than 0.00001). Therefore, diversification can widen the filter bubbles around the users by recommending news from more diverse topics.

\begin{table*}[h]
\centering
  \caption{Pearson correlation between diversity (\textit{ILD@10}) and topic coverage (CT@10)}
    \label{tab:prearson} 
   
  \begin{tabular}{lcccc}
    \toprule
    & Roularta&Kwestie&Globo.com&Adressa\\
    \midrule
    R &0.9594&0.9541&0.9251&0.8411\\
    % N &36&36&36& 36\\
    P-value&\(<.00001\)&\(<.00001\)&\(<.00001\)&\(<.00001\)\\
  \bottomrule
\end{tabular}
\end{table*}

The proposed diversification approaches have some limitations. Users' perception of diversity may differ from the diversity that we aim to enhance in this study ~\cite{ekstrand2014user}. A user study should be performed to verify whether users perceive the enhanced diversity in the recommendations lists. Moreover, in this study we only focused on content diversity but ideally a diversity-aware news recommender should also recommend diverse news w.r.t. sentiment and polarity. 

\section{Conclusion}
\label{sec6}
The main contribution of this study is to make neighborhood-based session-based recommender systems (SBRSs) diversity-aware. In news aggregator websites, focusing only on predictive performance of the recommender, can tighten the filter bubbles around the users and can intensify polarization and fragmentation among them. Diversification is a way to address these issues in news recommenders. We proposed scenarios to diversify the recommendation lists generated by these SBRSs. According to the results, all the scenarios improve diversity in all news datasets. The choice of the most effective approach depends on the method and the dataset. In all cases we can find a trade-off between diversity and accuracy. The decision as to which diversification approach should be applied can be seen as a hyperparameter and selected based on the validation set. The diversification approach which has the highest \textit{RR-ILD@10} in the validation phase should be applied on the test set.  Moreover, diversification addresses the filter bubble phenomenon by increasing the number of unique news topics in the recommendation lists. 

For future extension, we will assess the possibility of enhancing the diversity of model-based SBRSs such as GRU4REC~\cite{hidasi2018recurrent} and CHAMELEON~\cite{gabriel2019contextual}. In the loss functions of these model-based methods, regularization terms that penalize similar contents should be applied. Moreover, we will apply the proposed scenarios on other domains such as music and e-commerce recommenders. In these domains there are other types of metadata such as lyrics, genres, artists, item descriptions or a hierarchy of item categories that should be used to diversify recommendations. Finally, we will investigate how the performance of neighborhood-based SBRSs can be enhanced in other performance criteria such as fairness~\cite{GHARAHIGHEHI2021102663}, coverage~\cite{hyper_news_multi} and serendipity.

\begin{acknowledgements}
This work was executed within the imec.icon project NewsButler, a research project bringing together academic researchers (KU Leuven, VUB) and industry partners (Roularta Media Group, Bothrs and ML6). The NewsButler project is co-financed by imec and receives project support from Flanders Innovation \& Entrepreneurship (project nr. HBC.2017.0628). The authors also acknowledge support from the Flemish Government (AI Research Program).

“This is a post-peer-review, pre-copy edit version of an article published in \textit{Personal and Ubiquitous Computing}. The final authenticated version is available online at:

http://dx.doi.org/10.1007/s00779-021-01606-4”.
\end{acknowledgements}

\section*{Conflict of interest}
The authors declare that they have no conflict of interest.

% \bibliographystyle{spmpsci}      % mathematics and physical sciences
% \bibliographystyle{spphys}       % APS-like style for physics
% \bibliography{main}   % name your BibTeX data base

\end{document}